\documentclass[a4paper,notitlepage,eqsecnum,nofootinbib]{revtex4-1}
\usepackage{slashed,bbold,bm,amsmath,amssymb,mathtools,wrapfig,epsfig} 
\usepackage[usenames]{color}
\usepackage{hyperref}
\usepackage{tcolorbox}

\hypersetup{
    colorlinks=true,       
    linkcolor=red,          
    citecolor=blue,        
    filecolor=green,      
    urlcolor=magenta           
}

\newcommand{\crt}{\\[2mm]}
\newcommand{\nn}{\nonumber}

\newcommand{\bs}[1]{\boldsymbol{#1}}

\newcommand{\ie}{{\it i.e.}}

\newcommand{\eg}{{\it e.g.}}

\newcommand{\as}{\alpha_s}

\newcommand{\La}{\Lambda}
\newcommand{\la}{\lambda}

\newcommand{\vphi}{\varphi}
\newcommand{\order}[1]{${\cal O}\left(#1 \right)$}
\newcommand{\morder}[1]{{\cal O}\left(#1 \right)}
\newcommand{\eq}[1]{(\ref{#1})}
\newcommand{\fig}[1]{Fig.~\ref{#1}}

\newcommand{\inv}[1]{\frac{1}{#1}}
\newcommand{\halft}{{\textstyle \frac{1}{2}}}
\newcommand{\quart}{{\textstyle \frac{1}{4}}}

\newcommand{\intt}{{\textstyle \int}}
\newcommand{\ket}[1]{\left\vert{#1}\right\rangle}
\newcommand{\ketb}[1]{\vert{#1}\rangle}
\newcommand{\bra}[1]{\left\langle{#1}\right\vert}
\newcommand{\brab}[1]{\langle{#1}\vert}

\newcommand{\com}[2]{\left[{#1},{#2}\right]}
\newcommand{\comb}[2]{\big[{#1},{#2}\big]}
\newcommand{\acom}[2]{\left\{{#1},{#2}\right\}}
\newcommand{\acomb}[2]{\big\{{#1},{#2}\big\}}
\newcommand{\tr}{\mathrm{Tr}\,}

\newcommand{\mJ}{\mathcal{J}}

\newcommand{\mL}{\mathcal{L}}
\newcommand{\mM}{\mathcal{M}}

\newcommand{\mS}{\mathcal{S}}

\newcommand{\xv}{{\bs{x}}}

\newcommand{\yv}{{\bs{y}}}
\newcommand{\zv}{{\bs{z}}}

\newcommand{\pv}{{\bs{p}}}

\newcommand{\Av}{{\bs{A}}}

\newcommand{\Ev}{{\bs{E}}}

\newcommand{\Jv}{{\bs{J}}}

\newcommand{\Lv}{{\bs{L}}}
\newcommand{\Pv}{{\bs{P}}}

\newcommand{\Sv}{{\bs{S}}}

\newcommand{\gv}{\bs{\gamma}}

\newcommand{\gz}{\gamma^0}

\newcommand{\gf}{\gamma_5}

\newcommand{\qb}{{\bar{q}}}

\newcommand{\rar}{\rightarrow}
\newcommand{\lar}{\leftarrow}

\newcommand{\rham}[1]{{\overset{\rar}{H}}\strut_{\hspace{-.5mm}{#1}}}
\newcommand{\lham}[1]{{\overset{\lar}{H}}\strut_{\hspace{-.5mm}{#1}}}

\newcommand{\nv}{\bs{\nabla}}

\newcommand{\rnab}{{\overset{\rar}{\nv}}\strut}
\newcommand{\lnab}{{\overset{\lar}{\nv}}\strut}

\newcommand{\alv}{{\bs{\alpha}}}

\newcommand{\dt}{\partial_t}

\def\XXint#1#2#3{{\setbox0=\hbox{$#1{#2#3}{\int}$}
     \vcenter{\hbox{$#2#3$}}\kern-.5\wd0}}

\begin{document}

\title{Principles and Possibilities for Bound States in Gauge Theory}

\author{Paul Hoyer}
\affiliation{ \vspace{1mm} Department of Physics, POB 64, FIN-00014 University of Helsinki, Finland}
\email{paul.hoyer@helsinki.fi} 

\begin{abstract}  

Bound states differ from scattering yet are not covered in textbooks on Quantum Field Theory. I discuss a perturbative method for QED and QCD based on canonical quantization. Fully fixing temporal gauge  $A^0(t,\xv)=0$ imposes Gauss' law on physical states. As pointed out by Dirac, this implies that electron states include a longitudinal gauge field $\Av_L$, which determines the instantaneous bound state potential. The situation is analogous for quarks and gluons in QCD. An instantaneous confining potential arises for color singlet $q\qb$ states when a non-vanishing boundary condition on $\Av_L^a(\xv\to\infty)$ is specified in Gauss' constraint. As suggested by Gribov, $\as(Q^2)$ may freeze at a perturbative value when the confining potential dominates. Hadrons can then be calculated perturbatively. At vanishing quark mass there is a $j^{PC}=0^{++}$ state with zero energy which can mix with the perturbative vacuum, giving rise to a spontaneous breaking of chiral symmetry.

\end{abstract}

\maketitle

\vspace{-.5cm}
\tableofcontents

\parindent 0cm

\vspace{-.5cm}
\section{Introduction} \label{secI}

\vspace{-.3cm}
Gauge and Poincar\'e invariance are exact symmetries of the QED and QCD actions. The way that Poincar\'e covariance is realized depends on the gauge. The gauge field action,
\begin{align} \label{1.1}
-\quart\int dt\,d\xv F_{\mu\nu}F^{\mu\nu} = \halft\int dt\,d\xv\big[(\dt\Av+\nv A^0)^2-(\nv\times\Av_T)^2\big] +\morder{A^3,A^4}
\end{align}
has no $\dt A^0$ nor $\nv\cdot\Av_L$ terms. Hence $A^0$ and the longitudinal gauge field $\Av_L$ do not propagate in space-time (and ordinary light has only two transverse polarization components). All four components of the gauge field can be made to propagate by adding a Feynman gauge-fixing term to the action,
\begin{align} \label{1.2}
\mathcal{S}_{GF} = -\halft\int dt\,d\xv (\partial_\mu A^\mu)^2 = -\halft\int dt\,d\xv(\dt A^0+\nv\cdot\Av)^2
\end{align}
This gauge choice makes Feynman diagrams explicitly Poincar\'e symmetric.

Bound states are eigenstates of the Hamiltonian, \ie, stationary in time. This definition distinguishes time from space. Frame dependence is defined by the generators of Poincar\'e transformations, some of which include interactions. Binding energies may be calculated in the rest frame \cite{Adkins:2022omi}, with the CM momentum $\Pv$ dependence given by $E=\sqrt{\Pv^2+M^2}$. The $\Pv$-dependence of equal-time wave functions is on the other hand essentially unknown. This is surprising, also in view of the popular interest in classical Lorentz contraction.

The bound state method I discuss here differs in many respects from previous approaches, and builds on principles discussed by Dirac in 1955 \cite{Dirac:1955uv}. I use canonical quantization and states defined at equal ordinary time. The temporal gauge condition $A^0(t,\xv)=0$ gives rise to instantaneous interactions, since it is imposed at each instant of time over all space. Even particles in relativistic motion can interact instantaneously in gauge theories. This enables binding without the retardation caused by propagating (transverse) gauge bosons. 

Canonical quantization imposes equal-time commutation relations between fields and their conjugates. The $A^0$ field lacks a conjugate field due to the absence of a $\dt A^0$ term in gauge theory actions \eq{1.1}. This issue is avoided in temporal gauge, where both $A^0$ and its conjugate vanish. The Hamiltonian and other Poincar\'e generators are then local functionals of the fields. I am not aware of previous uses of $A^0=0$ gauge for bound states.

I recall the relevant features of temporal gauge \cite{Willemsen:1977fr,Bjorken:1979hv,Christ:1980ku,Leibbrandt:1987qv,Strocchi:2013awa} in Section \ref{secII}. Complete gauge fixing requires to impose Gauss' law on physical states. As emphasized by Dirac \cite{Dirac:1955uv}, physical charged particle states (\eg, electrons) include an instantaneous longitudinal gauge field $\Av_L$. This implies an electric field $\Ev_L$, whose energy is the bound state potential (the Coulomb potential $-\alpha/r$ of Positronium). 

\begin{figure}[h] \centering
\includegraphics[width=.8\columnwidth]{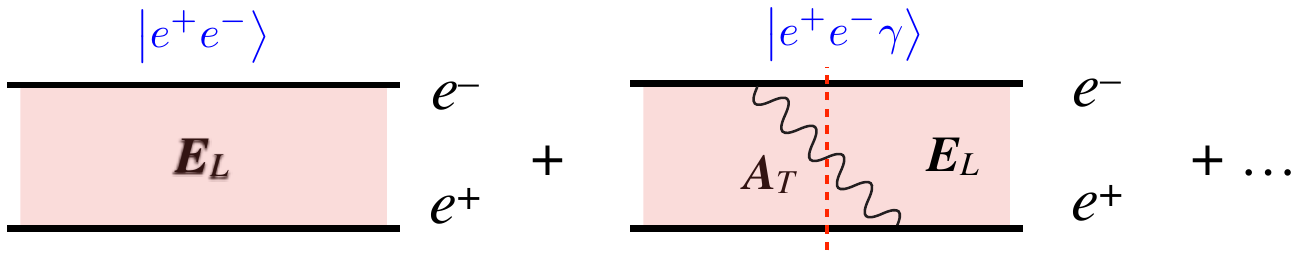}
\caption{Sketch of the perturbative expansion for the $\ket{e^+e^-}$ Fock state of Positronium. The longitudinal electric field $\Ev_L$ of the constituents generates an instantaneous potential (pink shading). Transverse photons ($\Av_T$, wavy line in right diagram) and $e^+e^-$ pairs are added perturbatively to the valence approximation. In both diagrams time is increasing to the right.
\label{f1}}
\end{figure}

The perturbative expansion for the $\ket{e^+e^-}$ Fock component of a Positronium atom is sketched in \fig{f1}. In a first (valence) approximation (left) the $e^+e^-$ pair interacts only through the instantaneous Coulomb potential due to $\Ev_L$. This determines the $e^+e^-$ wave function as a non-polynomial (exponential) function of $\alpha$.  Applying the  Hamiltonian (twice) to this state creates an intermediate state with a transverse photon (right, wavy line). Fock states with any number of transverse photons and $e^+e^-$ pairs are generated by corresponding powers of the Hamiltonian (\ie, of $\alpha$). The expansion is possible in all reference frames. This was checked \cite{Hoyer:2021adf} for Positronium, by including the $\ket{e^+e^-\gamma}$ state (shown at dashed red line), which contributes to the binding energy at leading order for atoms in motion \cite{Jarvinen:2004pi}.

An expansion like \fig{f1} is motivated also for hadrons in QCD. Hadrons are classified according to their valence quark content ($q\qb,\,qqq,\ldots$). This indicates that their higher Fock components $q\qb g,\ldots$ can be treated perturbatively. The increase of $\as(Q^2)$ towards smaller $Q^2$ should stop at the scale given by hadron radii. The ``frozen'' value $\as(0)$ can be small enough to allow a perturbative expansion, as advocated in \cite{Gribov:1999ui,Dokshitzer:2004ie}. 

The Schr\"odinger equation with the Cornell potential \cite{Eichten:1979ms,Eichten:2007qx}
\begin{align} \label{1.3}
V(r) = V'r-\frac{4}{3}\frac{\as}{r} \ \ \ \text{with}\ \ V' \simeq 0.18\ \text{GeV}^2, \ \ \as \simeq 0.39
\end{align}
gives a good description of heavy quarkonia \cite{Brambilla:2010cs}. The potential was determined from data, and later found to agree with (quenched) lattice QCD \cite{Bali:2000gf}. 

In the present approach confinement arises from a boundary condition on Gauss' law for the longitudinal electric field $\Ev_L^a$ (Sect. \ref{secV}). The hadron scale $V'$, which is not in the QCD action, is thus introduced through the definition of temporal gauge.
Confinement allows to consider QCD bound states at \order{\as^0}, keeping only the $V'r$ term in \eq{1.3}. At $\as=0$ the perturbative corrections (transverse gluons, $q\qb$ pairs) are absent, \ie, only the left diagram of \fig{f1} (adapted to QCD) contributes. The \order{\as^0} $q\qb$ states have features in common with physical hadrons, including linear Regge trajectories \cite{Hoyer:2021adf}. Boost covariance is ensured by the generator algebra at each order of $\as$.

\section{Temporal gauge} \label{secII}

Temporal gauge, $A^0(t,\xv) =0$, is reviewed in \cite{Willemsen:1977fr,Bjorken:1979hv,Christ:1980ku,Leibbrandt:1987qv,Strocchi:2013awa}, see also \cite{Hoyer:2021adf}. Here I recall some of its features in QED. The action
\begin{align} \label{2.1}
\mS_{QED} &= \int d^4x\big[-\quart F_{\mu\nu}F^{\mu\nu}+\bar\psi(i\slashed{\partial}-m-e\slashed{A})\psi\big]
& F_{\mu\nu} = \partial_\mu A_\nu-\partial_\nu A_\mu
\end{align}
determines the electric field $E^i=F^{i0}=-\partial_0 A^i$ to be conjugate to $A_i \ (i=1,2,3)$, and $i\psi^\dag$ to be conjugate to $\psi$. This gives the non-vanishing canonical commutation relations, 
\begin{align} \label{2.2}
\com{E^i(t,\xv)}{A^j(t,\yv)} &= i\,\delta^{ij}\delta(\xv-\yv)  & \acom{\psi^\dag_\alpha(t,\xv)}{\psi_\beta(t,\yv)} = \delta_{\alpha\beta}\,\delta(\xv-\yv)
\end{align}

The temporal gauge condition $A^0=0$ does not fully fix the gauge. It allows time independent gauge transformations, which are generated by Gauss' operator $G(t,\xv)$,
\begin{align} \label{2.3}
G(t,\xv) \equiv \frac{\delta\mS_{QED}}{\delta{A^0(t,\xv)}} = \nv\cdot \Ev(t,\xv)-e\psi^\dag\psi(t,\xv)
\end{align}
The longitudinal and transverse parts of the electric field, $\Ev(t,\xv)= \Ev_L+\Ev_T$, satisfy $\nv\times\Ev_L=0$ and $\nv\cdot\Ev_T=0$. The gauge invariance of physical states is ensured by requiring them to be annihilated by $G(t,\xv)$,
\begin{align} \label{2.4}
G(t,\xv)\ket{phys} = \big[\nv\cdot \Ev_L(t,\xv)-e\psi^\dag\psi(t,\xv)\big]\ket{phys} = 0 \hspace{2cm} G(t,\xv)\ket{0}=0
\end{align}
If this condition holds at $t=0$ it is valid at all times since $G(t,\xv)$ commutes with the Hamiltonian, $\com{G(t,\xv)}{H(t)}=0$.

To illustrate, consider a bare electron state created by $b^\dag_{\pv,\la}$ at $t=0$,
\begin{align} \label{2.5}
\ket{e^-;\pv,\la}_0 \equiv b^\dag_{\pv,\la}\ket{0} = \int d\zv\,\psi^\dag(t=0,\zv)u(\pv,\la)e^{i\pv\cdot\zv}\ket{0} 
\end{align}
This state is unphysical in temporal gauge since $G(\xv)\ket{e^-;\pv,\la}_0 \neq 0$. A fully gauge fixed, physical electron state at $t=0$ includes a longitudinal photon field $\Av_L$ \cite{Dirac:1955uv}\footnote{Dirac's ``gauge invariant'' formulation of QED appears to be equivalent to fully fixed temporal gauge. The field denoted $\psi^*$ in Eq. [31] of \cite{Dirac:1955uv} corresponds to $\psi^\dag W$ in \eq{2.6}.},
\begin{align} \label{2.6}
\ket{e^-;\pv,\la} &\equiv \int d\zv\,\psi^\dag(\zv)\,W[\Av_L;\zv]u(\pv,\la)e^{i\pv\cdot\zv}\ket{0} \crt
W[\Av_L;\zv] &= \exp\big[i\int d\yv\, \Av_L(\yv)\cdot\nv_y V_I(\yv,\zv)\big] \nn
\end{align} 
The derivative $\nv_y$ of the c-numbered function $V_I(\yv,\zv)$ is included for convenience. The field ordering in the exponential $W[\Av_L;\zv]$ is irrelevant since $\comb{A^j(t,\yv)}{A^k(t,\yv')}=0$.

The electron field contribution to $G(t=0,\xv)$ \eq{2.4} is determined by \eq{2.2} in the integrand of the state \eq{2.6},
\begin{align} \label{2.7}
\com{e\,\psi^\dag\psi(\xv)}{\psi_\alpha^\dag(\zv)} = e\,\psi_\alpha^\dag(\xv)\,\delta(\zv-\xv)
\end{align}
According to \eq{2.4} this should be equal to the contribution of $\nv\cdot\Ev(\xv)$. Since
\begin{align} \label{2.8}
\big[\nv_x\cdot \Ev_L(\xv),W[\Av_L;\zv]\big] = W[\Av_L;\zv]\, i^2\nv_x\cdot\int d\yv\,\delta(\xv-\yv)\nv_y V_I(\yv,\zv) = -W[\Av_L;\zv] \nv_x^2 V_I(\xv,\zv)
\end{align}
$V_I(\xv,\zv)$ should be the Coulomb potential, and $-\nv_x V_I(\xv,\zv)$ the electric field at $\xv$ of a charge at $\zv$,
\begin{align} \label{2.9}
\nv_x^2 V_I(\xv,\zv) =-e\,\delta(\xv-\zv) \hspace{1.5cm}i.e., \hspace{1.5cm}  V_I(\xv,\zv) = \frac{e}{4\pi|\xv-\zv|}
\end{align}

Gauss constraint \eq{2.4} allows to express the longitudinal electric field acting on any physical QED state in terms of the electron field,
\begin{align} \label{2.10}
\Ev_L(t,\xv)\ket{phys} = -\nv_\xv \int d\yv\,\frac{e}{4\pi|\xv-\yv|}\,\psi^\dag\psi(t,\yv)\ket{phys}
\hspace{2cm} \Ev_L(t,\xv)\ket{0} = 0
\end{align}
The latter condition \cite{Dirac:1955uv} implies that Gauss' constraint does not involve particle creation. As we shall see below, this explains why the strong confining field of QCD does not create quark pairs and gluons, allowing hadrons to be approximated by their valence Fock state.

The Hamiltonian in temporal gauge is, with $\alv\equiv\gz\gv$,
\begin{align} \label{2.11}
H &= \int d\xv\big[\halft (\Ev_L^2+\Ev_T^2) +\halft (\nv\times\Av_T)^2+\psi^\dag(-i\alv\cdot\rnab+m\gz-e\,\alv\cdot\Av)\psi\big] 
\end{align}
Using \eq{2.10},
\begin{align} \label{2.12}
\halft\int d\xv\,\Ev_L^2(\xv)\ket{phys} &= \halft\int d\xv d\yv d\zv\,\Big[\nv_x\,\frac{e}{4\pi|\xv-\yv|}\,\psi^\dag\psi(\yv)\Big]\Big[\nv_x\,\frac{e}{4\pi|\xv-\zv|}\,\psi^\dag\psi(\zv)\Big]\ket{phys} \nn\crt
&= \halft\int d\yv d\zv\,\frac{e^2}{4\pi|\yv-\zv|}\big[\psi^\dag\psi(\yv)\big]\big[\psi^\dag\psi(\zv)\big]\ket{phys}
\end{align}

The kinetic energy of a physical electron with momentum $\pv$ is $E_p = \sqrt{\pv^2+m^2}$,
\begin{align} \label{2.13}
\int d\zv\,\psi^\dag(-i\alv\cdot\rnab+m\gz)\psi \ket{e^-;\pv,\la} &=
\int d\xv\,\psi^\dag(\xv)W[\Av_L;\xv](i\alv\cdot\lnab_x+\gz m)u(\pv,\la)e^{i\pv\cdot\xv}\ket{0} \nn\crt
&= \int d\xv\,\psi^\dag(\xv)W[\Av_L;\xv](\alv\cdot\pv+\gz m)u(\pv,\la)e^{i\pv\cdot\xv}\ket{0}
= E_p\ket{e^-;\pv,\la} 
\end{align}
The derivative $\lnab_x$ acts on $\psi^\dag(\xv)\,W[\Av_L;\xv]$ since for physical states any shift in the position of the electron implies a shift in $W[\Av_L;\xv]$. In effect, the operator $\psi^\dag(\xv)\,W[\Av_L;\xv]$\ \ ($\equiv \psi^*(\xv)$ in \cite{Dirac:1955uv}) is defined by a single position $\xv$.

\section{Positronium in Temporal gauge} \label{secIII}

A physical $e^+e^-$ Fock state in the rest frame may at $t=0$ be defined as,
\begin{align}
\ket{e^+e^-}&=\int d\xv_1 d\xv_2\,\bar\psi_\alpha(\xv_1)\Phi_{\alpha\beta}(\xv_1-\xv_2) W[\Av_L;\xv_1,\xv_2]\psi_\beta(\xv_2)\ket{0} \label{3.1} \crt
W[\Av_L;\xv_1,\xv_2] &\equiv \exp\Big[\frac{ie}{4\pi}\int d\yv\,\Av_L(\yv)\cdot\nv_y
\Big(\inv{|\yv-\xv_1|}-\inv{|\yv-\xv_2|}\Big)\Big] \label{3.2}
\end{align}
The electron with its $\Av_L$ field is created at $\xv_1$ and the positron at $\xv_2$. Their distribution in space is governed by the c-numbered $(4\times 4)$ wave function $\Phi_{\alpha\beta}$, which in the rest frame depends only on the separation $\xv_1-\xv_2$.

Consider now the longitudinal electric field energy of $\ket{e^+e^-}$, determined by the Hamiltonian \eq{2.11} 
\begin{align} \label{3.3}
\com{E_L^j(\xv)}{W[\Av_L;\xv_1,\xv_2]} &= W[\Av_L;\xv_1,\xv_2]\,\frac{i^2e}{4\pi}\,\nabla_{x,j}\Big(\inv{|\xv-\xv_1|}-\inv{|\xv-\xv_2|}\Big) \nn\crt
\com{E_L^j(\xv)}{\com{E_L^j(\xv)}{W[\Av_L;\xv_1,\xv_2]}} &=W[\Av_L;\xv_1,\xv_2]\,\frac{i^4e^2}{(4\pi)^2}\Big[\nv_x\Big(\inv{|\xv-\xv_1|}-\inv{|\xv-\xv_2|}\Big)\Big]^2 \nn\crt
\halft\int d\xv\,\com{\Ev_L^2(\xv)}{W[\Av_L;\xv_1,\xv_2]} &=W[\Av_L;\xv_1,\xv_2]\,\frac{-e^2}{2(4\pi)^2}\int d^3\xv\Big(\inv{|\xv-\xv_1|}-\inv{|\xv-\xv_2|}\Big) \nv^2_x\Big(\inv{|\xv-\xv_1|}-\inv{|\xv-\xv_2|}\Big) \nn\crt
&= -\frac{\alpha}{|\xv_1-\xv_2|}\, W[\Av_L;\xv_1,\xv_2] \equiv V(|\xv_1-\xv_2|)\, W[\Av_L;\xv_1,\xv_2]
\end{align}
where $\alpha=e^2/4\pi$ and I subtracted the electron and positron self-energies $\propto 1/0$. The same result is obtained using \eq{2.12} and the commutator \eq{2.7} of the electron field. The Coulomb potential $V(\xv)=-\alpha/|\xv|$ equals the energy of the instantaneous $\Av_L$ field. The interaction term $\int d\xv\,\psi^\dag(-e\,\alv\cdot\Av)\psi$ of the Hamiltonian \eq{2.11} contributes spin and annihilation corrections of \order{\alpha^4} to the Positronium energy in the rest frame \cite{Adkins:2022omi,Hoyer:2021adf}. 

The non-relativistic limit of the $e^\pm$ kinetic energies \eq{2.13} are, 
\begin{align} \label{3.4}
\int d\xv\,\comb{\psi^\dag(-i\alv\cdot\rnab+m\gz)\psi}{\bar\psi(\xv_1)} &\to \bar\psi(\xv_1)\sqrt{-\lnab_1^2+m^2}\ \simeq \bar\psi(\xv_1)\Big(m- \frac{\lnab_1^2}{2m}\Big)\nn\crt
\int d\xv\,\comb{\psi^\dag(-i\alv\cdot\rnab+m\gz)\psi}{\psi(\xv_2)} &\to \sqrt{-\rnab_2^2+m^2}\ \psi(\xv_2)\simeq \Big(m- \frac{\rnab_2^2}{2m}\Big) \psi(\xv_2)
\end{align}
Altogether at \order{\alpha^2}, 
\begin{align} \label{3.5}
H\ket{e^+e^-} = \int d\xv_1 d\xv_2\,\bar\psi(\xv_1)W\Big[\Big(2m- \frac{\rnab_1^2}{2m}- \frac{\rnab_2^2}{2m}-\frac{\alpha}{|\xv_1-\xv_2|}\,\Big)\Phi(\xv_1-\xv_2)\Big]\psi(\xv_2)\ket{0}= (2m+E_b)\ket{e^+e^-}
\end{align}
where $E_b$ is the binding energy. This implies the Schr\"odinger equation for the rest frame Positronium wave function,
\begin{align} \label{3.6}
\Big(- \frac{\nv^2}{m}-\frac{\alpha}{|\xv|}\,\Big)\Phi(\xv) = E_b\,\Phi(\xv)
\end{align}

\section{QCD Meson} \label{secIV}

\subsubsection{Transverse Gauss operator} \label{secIV.1}

The QCD action is, with color indices denoted $A,B,C,\ldots$ for quarks and $a,b,c,\ldots$ for gluons,
\begin{align} \label{4.1}
\mS_{QCD} &= \int d^4x\big[-\quart F_{\mu\nu}^aF^{\mu\nu}_a+\bar\psi(i\slashed{\partial}-m-g\slashed{A}_a T^a)\psi\big]
& F_{\mu\nu}^a = \partial_\mu A_\nu^a-\partial_\nu A_\mu^a-gf_{abc}A_\mu^bA_\nu^c
\end{align}
The electric field in temporal gauge is $E_a^i=F_a^{i0}=-\dt A_a^i$ is conjugate to $A_i^a=-A_a^i$, with the commutation relations
\begin{align} \label{4.2}
\com{E_a^i(t,\xv)}{A_b^j(t,\yv)} &= i\delta_{ab}\delta^{ij}\delta(\xv-\yv)  & \acom{\psi^{A\,\dag}_\alpha(t,\xv)}{\psi_\beta^B(t,\yv)} = \delta^{AB}\delta_{\alpha\beta}\,\delta(\xv-\yv)
\end{align}
The Hamiltonian is in temporal gauge
\begin{align} \label{4.3}
H_{QCD} &= \int d\xv\big[E_a^i\dt A_i^a+i\psi_A^\dag\dt\psi_A-\mL_{QCD}\big]
= \int d\xv\big[\halft \Ev_a^2 +\quart (F_a^{ij})^2+\psi^\dag(-i\alv\cdot\rnab+m\gz-g\alv\cdot \Av_a T^a)\psi\big]
\end{align}

Gauss' operator
\begin{align} \label{4.4}
G_a(x) \equiv \frac{\delta\mS_{QCD}}{\delta{A_a^0(x)}}
=\nv\cdot \Ev_L^{a}(x)+g f_{abc}\Av^b \cdot\Ev^c(x) -g\psi^\dag T^a\psi(x)
\end{align}
generates gauge transformations $U_G$ that preserve the temporal gauge condition $A_a^0(t,\xv)=0$. With $\dt\La_b(\yv) =0$,
\begin{align} \label{4.5}
U_G(t) = 1+ i\int d\yv\,G_b(t,\yv)\La_b(\yv) + \morder{\La^2}
\end{align}
The space components of the gauge field transform as
\begin{align} \label{4.6}
\delta A_a^k(t,\xv) &= U_G(t)\, A^k_a(t,\xv)\,U_G^\dag(t) - A^k_a(t,\xv)
= i\int d\yv\,\com{G_b(t,\yv)}{A_a^k(t,\xv)}\La_b(\yv) \nn\crt
&= -i\int d\yv\,\com{E^j_b(t,\yv)}{A^k_a(t,\xv)}\partial_j^y\La_b(\yv)
 +i\int d\yv\,\com{gf_{bcd}A_c^j E^j_d(t,\yv)}{A_a^k(t,\xv)}\La_b(\yv) \nn\crt
 &= \partial_k^x\La_a(\xv) + gf_{abc}A_b^k(t,\xv)\La_c(\xv)  + \morder{\La^2}
\end{align}
Longitudinal ($\nv\times \Av_{L,a} =0$) gauge fields get a transverse component in the $U_G$ transformation,
\begin{align} \label{4.7}
\nv \times \delta \Av_{L,a} = -gf_{abc}\, \Av_{L,b}\times \nv \La_c \neq 0
\end{align}

The physical states of temporal gauge are defined as in \eq{2.6}. Each charge is the source of an instantaneous \textit{longitudinal} gauge field. In effect, this implies gauge fixing beyond $A_a^0=0$. Gauge transformations $U_T$ which maintain both $A_a^0=0$ and the longitudinal nature of $\Av_{L,a}$ are generated by the ``transverse Gauss' operator'' $G_{T,a}(\xv)$,
\begin{align} \label{4.8}
U_T &= 1+i\int d\yv\, G_{T,b}(\yv)\,\La_b(\yv) + \morder{\La^2} \hspace{2cm} G_{T,a} \equiv \nv\cdot \Ev_{a} + gf_{abc}\Av_{T,b}\cdot\Ev_{T,c} - g\psi^\dag T^a\psi \crt
& \delta{\Av_{L,a}}(\xv) =  i\int d\yv\,\com{G_{T,b}(\yv)}{\Av_{L,a}(\xv)}\La(\yv) =\nv\La_a(\xv)  \nn
\end{align}

\subsubsection{Coulomb potential} \label{secIV.2}

In analogy to QED \eq{3.1} I define $q\qb$ Fock states at rest as
\begin{align}
\ket{q\qb} &\equiv \int d\xv_1 d\xv_2\,\bar\psi_A(\xv_1)\Phi(\xv_1-\xv_2)W_{AB}[\Av_L;\xv_1,\xv_2] \psi_B(\xv_2)\ket{0}  \label{4.9}\crt
W[\Av_L;\xv_1,\xv_2] &\equiv \exp\Big[\frac{ig}{4\pi}\int_{\xv_2}^{\xv_1} d\yv\,T^a\Av_L^a(\yv)\cdot\nv_y \Big(\inv{|\yv-\xv_1|}-\inv{|\yv-\xv_2|}\Big)\Big]  \label{4.10}
\end{align}
The $\yv$-integrals follow the field lines starting at $\xv_2$ and ending at $\xv_1$.
Full gauge fixing requires that the state $\ket{q\qb}$ \eq{4.9} is invariant under the gauge transformations generated by $G_{T,a}(\xv)$, \ie, $G_{T,a}(\xv)\ket{q\qb} = 0$. The gauge functional $W[\Av_L;\xv_1,\xv_2]$ contributes, 
\begin{align} \label{4.11}
\big[\nv_x\cdot\Ev_L^a(\xv),W[\Av_L;\xv_1,\xv_2]\big] &= W[\Av_L;\xv_1,\xv_2]\frac{i^2g}{4\pi}\,T^a\nv_x^2\Big(\inv{|\xv-\xv_1|}-\inv{|\xv-\xv_2|}\Big) \nn\crt
&= gT^a\,W[\Av_L;\xv_1,\xv_2]\delta(\xv-\xv_1)-\delta(\xv-\xv_2)W[\Av_L;\xv_1,\xv_2]gT^a 
\end{align}
In the first line $T^a$ is understood to appear at $\yv=\xv$ within the exponential $W$ \eq{4.10}. The only non-vanishing contributions are at the position $\xv = \xv_1\ (\xv_2)$ of the $q\ (\qb)$. Similarly for the quark contributions,
\begin{align} \label{4.12}
g\com{\psi^\dag(\xv)T^a\psi(\xv)}{\bar\psi_\alpha(\xv_1)W\psi_\beta(\xv_2)} &= \bar\psi_\alpha(\xv_1)gT^a\,\delta(\xv-\xv_1)W\psi_\beta(\xv_2)- \bar\psi_\alpha(\xv_1)W\delta(\xv-\xv_2)gT^a\,\psi_\beta(\xv_2) 
\end{align}
Together with \eq{4.11} this ensures $G_{T,a}(\xv)\ket{q\qb} = 0$. As expected, the longitudinal electric field $\Ev_{L}^{a}(\xv)$ is sourced by the transverse gluon and quark constituents of the state. In terms of the adjoint generator $(t^a)_{bc} = -if_{abc}$,
\begin{align} \label{4.13}
\nv\cdot \Ev_{L}^{a}(\xv)\ket{q\qb} = g\big[- i\,A_T^j\, t^a E_T^j(\xv)+\psi^\dag T^a\psi(\xv)\big]\ket{q\qb}
\end{align}

The energy of the longitudinal field is obtained similarly as in \eq{3.3},
\begin{align} \label{4.14}
\com{E_{L,a}^j(\xv)}{W[\Av_L;\xv_1,\xv_2]} &= W[\Av_L;\xv_1,\xv_2]\frac{i^2g}{4\pi}\,T^a\nabla_{x,j}\Big(\inv{|\xv-\xv_1|}-\inv{|\xv-\xv_2|}\Big) \nn\crt
\com{E_{L,a}^j(\xv)}{\com{E_{L,a}^j(\xv)}{W[\Av_L;\xv_1,\xv_2]}} &=W[\Av_L;\xv_1,\xv_2]\frac{i^4g^2}{(4\pi)^2}\,T^aT^a\Big[\nv_x\Big(\inv{|\xv-\xv_1|}-\inv{|\xv-\xv_2|}\Big)\Big]^2 \nn\crt
\halft\int d\xv\com{\Ev_{L,a}^2(\xv)}{W[\Av_L;\xv_1,\xv_2]} &=W[\Av_L;\xv_1,\xv_2]\frac{-g^2C_F}{2(4\pi)^2}\int d\xv\Big(\inv{|\xv-\xv_1|}-\inv{|\xv-\xv_2|}\Big) \nv^2\Big(\inv{|\xv-\xv_1|}-\inv{|\xv-\xv_2|}\Big) \nn\crt
&= -C_F\frac{\as}{|\xv_1-\xv_2|}\, W[\Av_L;\xv_1,\xv_2] 
\end{align}
where $C_F=T^aT^a$. The infinite contributions $\propto\alpha_s/0$ of the quark self-energies were subtracted. On the first lines $T^a$ and $T^aT^a$ are at $\yv=\xv$ within $W[\Av_L;\xv_1,\xv_2]$ \eq{4.10}. Since $T^aT^a = C_F$ is proportional to the unit color matrix it may be moved out of $W[\Av_L;\xv_1,\xv_2]$, giving the well-known ``gluon exchange'' potential.

The kinetic energy of non-relativistic quarks is as for Positronium \eq{3.4}, 
\begin{align} \label{4.15}
\int d\xv\,\psi^\dag(-i\alv\cdot\rnab+m\gz)\psi\ket{q\qb}=
\int d\xv_1 d\xv_2\,\bar\psi(\xv_1)W[\Av;\xv_1,\xv_2]\Big[\Big(2m- \frac{\rnab_1^2}{2m}- \frac{\rnab_2^2}{2m}\Big)\Phi(\xv_1-\xv_2)\Big]\psi(\xv_2)\ket{0}
\end{align}
The bound state condition $H_{QCD}\ket{q\qb}=2m+E_b$ thus implies the Schr\"odinger equation for the wave function,
\begin{align} \label{4.16}
\Big(- \frac{\nv^2}{m}-C_F\frac{\as}{|\xv|}\,\Big)\Phi(\xv) = E_b\,\Phi(\xv)
\end{align}

\section{Confinement} \label{secV}

The Schr\"odinger equation \eq{4.16} lacks a confining potential. The $1/r$ potential corresponds to dimensional analysis. Any other $r$-dependence requires a confinement scale, such as $V'$ in the Cornell potential \eq{1.3}. The QCD action \eq{4.1} has no such scale. A scale does arise in regularizing UV divergent QCD loop integrals. The successful description of heavy quarkonia in terms of the Schr\"odinger equation with a confining potential suggests that the physical scale may appear before loop corrections. This could occur through a boundary condition.

\subsubsection{Gauss' constraint} \label{secV.1}

Let us reconsider the derivation of Sect. \ref{secIV} for an implicit assumption.
The gauge exponential $W$ \eq{4.10} was inspired by QED \eq{3.2}, where it ensures that the electric field vanishes far from the charge. Color confinement in QCD makes this requirement less compelling. The Cornell potential $\propto V'r$ \eq{1.3} increases with $r$, but is quenched by quark pair production at large $r$ (string breaking).

Each quark color component of the $q\qb$ Fock state \eq{4.9} sources an octet field $\Ev_L^a(\xv)$, which is determined by $W$ and binds the quarks of that color. For color singlet states the octet field  cancels in the sum over quark colors. In Positronium the electron is similarly bound by the monopole field of the positron (Coulomb potential $-\alpha/r$). An external observer measures the sum of the electron and positron fields, which is a dipole. The cancellation is incomplete in QED.

Defining temporal gauge requires to specify also the boundary conditions (homogeneous solutions) of Gauss' constraint $G_{T,a}(\xv)\ket{q\qb}=0$ \eq{4.13}. Adding $\kappa\,\yv\cdot(\xv_1-\xv_2)$ to the Coulomb potentials in the exponent of $W$,
\begin{align} \label{5.1}
W_\kappa[\Av_L;\xv_1,\xv_2] &\equiv \exp\Big[i\int_{\xv_2}^{\xv_1} d\yv\,T^a\Av_L^a(\yv)\cdot\nv_y \Big[\frac{g}{4\pi}\Big(\inv{|\yv-\xv_1|}-\inv{|\yv-\xv_2|}\Big) + \kappa\,\yv\cdot(\xv_1-\xv_2)\Big]
\end{align}
does not affect $\nv\cdot\Ev_L^a$ \eq{4.11} if $\kappa$ is independent of $\yv$. The longitudinal electric field of $\ket{q\qb}$ becomes, 
\begin{align} \label{5.2}
\com{\Ev_{L}^a(\xv)}{W_\kappa[\Av_L;\xv_1,\xv_2]} &= W_\kappa[\Av_L]\,i^2\,T^a\nv_{x}\Big[\frac{g}{4\pi}\Big(\inv{|\xv-\xv_1|}-\inv{|\xv-\xv_2|}\Big)+\kappa\,\xv\cdot(\xv_1-\xv_2)\Big] 
\end{align}
where $T^a$ is understood to be at $\yv=\xv$ within $W_\kappa[\Av_L] \equiv W_\kappa[\Av_L;\xv_1,\xv_2]$. The contribution $\propto \kappa$ is $\xv$-independent and aligned with the quark separation, which is consistent with space translation and rotation symmetry. There appears to be no other homogeneous solution of Gauss' constraint for $\ket{q\qb}$ that preserves these symmetries.

The electric field energy \eq{4.14} is now,
\begin{align} \label{5.3}
\halft\int d\xv\,&\com{\Ev_{L,a}^2(\xv)}{W_\kappa[\Av_L;\xv_1,\xv_2]} = W_\kappa[\Av_L]\,\halft\,C_F\int d\xv\,\Big[\frac{g}{4\pi}\nv_x\Big(\inv{|\xv-\xv_1|}-\inv{|\xv-\xv_2|}\Big)+\kappa(\xv_1-\xv_2)\Big]^2 \nn\crt
&= W_\kappa[\Av_L]\,C_F\Big[-\frac{\as}{|\xv_1-\xv_2|} -\frac{g\kappa}{4\pi}\int d\xv\Big(\frac{\xv-\xv_1}{|\xv-\xv_1|^3} -\frac{\xv-\xv_2}{|\xv-\xv_2|^3}\Big)\cdot(\xv_1-\xv_2) + \halft \kappa^2(\xv_1-\xv_2)^2\int d\xv\ \Big]
\end{align}
The integral multiplying $-g\kappa/4\pi$ depends only on $|\xv_1-\xv_2| \equiv 2\ell$,
\begin{align} \label{5.4}
I(\ell)\equiv \int d\xv\Big(\frac{\xv-\xv_1}{|\xv-\xv_1|^3}-\frac{\xv-\xv_2}{|\xv-\xv_2|^3}\Big)\cdot(\xv_1-\xv_2)
\end{align}
Choosing the $\xv$-coordinates such that $\xv_1=(0,0,\ell)$ and $\xv_2=(0,0,-\ell)$ are on the $z$-axis with $\ell>0$,
\begin{align} \label{5.5}
I(\ell) = 2\ell\int_{-\infty}^\infty dz \int d^2\xv_\perp\Big\{\frac{z-\ell}{[x_\perp^2+(z-\ell)^2]^{3/2}} -\frac{z+\ell}{[x_\perp^2+(z+\ell)^2]^{3/2}}\Big\}
\end{align}
The $\xv_\perp$ integral gives
\begin{align}
\int \frac{d^2\xv_\perp}{(x_\perp^2+a^2)^{3/2}} = 2\pi\int_0^\infty\frac{x_\perp dx_\perp}{(x_\perp^2+a^2)^{3/2}} =-2\pi\Big/_{\hspace{-1.5mm}0}^\infty \inv{(x_\perp^2+a^2)^{1/2}} = \frac{2\pi}{|\,a\,|} \nn
\end{align}
Hence
\begin{align} \label{5.6}
I(\ell) = 4\pi\ell\int_{-\infty}^\infty dz\, \Big(\frac{z-\ell}{|z-\ell|} -\frac{z+\ell}{|z+\ell|}\Big)
= 4\pi\ell\int_{-\ell}^\ell dz\,(-2) = -16\pi\ell^2 =-4\pi (\xv_1-\xv_2)^2
\end{align}
Using this in \eq{5.3} the energy of the longitudinal field energy in the $\ket{q\qb}$ bound state \eq{4.9} with $W \to W_\kappa$ becomes
\begin{align} \label{5.7}
\int d\xv\,\halft\big[\Ev_L^a(\xv)\big]^2\ket{q\qb} &= \int d\xv_1 d\xv_2\,\bar\psi(\xv_1)C_F\Big\{-\frac{\as}{|\xv_1-\xv_2|} + (\xv_1-\xv_2)^2\big[g\kappa + \halft \kappa^2\intt d\xv\big] \Big\}  \nn\crt
&\times\Phi(\xv_1-\xv_2)W_\kappa[\Av_L;\xv_1,\xv_2] \psi(\xv_2)\ket{0}
\end{align}

The \order{\kappa^2} term contributes an $\xv$-independent energy density $E_\La$ (analogous to the Bag Model density $B$ \cite{Chodos:1974je}), with total energy proportional to the volume of space. It is irrelevant only if it is universal, \ie, the same for all Fock states of all bound states. This determines the normalization $\kappa$ for any Fock state.

For the vacuum field energy density $E_\La$ to be independent of $\xv_1$ and $\xv_2$ we must have $\kappa \propto 1/|\xv_1-\xv_2|$ in \eq{5.7}. Defining $E_\La$ in terms of a universal constant $\La$,
\begin{align} \label{5.8}
E_\La = \frac{\La^4}{2C_F g^2} \hspace{3cm}  \kappa = \frac{\La^2}{C_Fg}\,\inv{|\xv_1-\xv_2|}
\end{align}
Subtracting the (infinite) field energy $E_\La\int d\xv$ in \eq{5.7} defines the potential $V(\xv)$, which with $V'=\La^2$ agrees with the Cornell potential \eq{1.3} for any quark mass,
\begin{align} \label{5.9}
\int d\xv\,\halft\big[\Ev_L^a(\xv)\big]^2\ket{q\qb} &= \int d\xv_1 d\xv_2\,\bar\psi(\xv_1)\Big[\La^2\,|\xv_1-\xv_2|-C_F\frac{\as}{|\xv_1-\xv_2|}
 \Big]\Phi(\xv_1-\xv_2)W_\kappa[\Av_L;\xv_1,\xv_2] \psi(\xv_2)\ket{0} \nn\crt
V(\xv) &= \La^2|\xv| - C_F\frac{\as}{|\xv|}
\end{align}
The scale $\La$ is independent of the coupling $\as$. We may set $\as=0$ and consider the \order{\as^0} states bound by the linear $V(\xv)=\La^2|\xv|$ potential. The longitudinal electric field does not create gluon or quark pair constituents, $\Ev_L^a\ket{0}=0$ \eq{2.10}. Without the \order{g} interaction terms in the Hamiltonian \eq{4.3} $\ket{q\qb}$ is thus a pure valence $q\qb$ state. The perturbative suppression of \order{\as} contributions may explain why physical hadrons, in spite of being strongly bound by the confining potential, are dominated by their valence Fock states.

\subsubsection{Bound state equation} \label{secV.2}

Consider again the rest frame $\ket{q\qb}$ bound state \eq{4.9} with mass $M$,
\begin{align} \label{5.10}
\ket{q\qb;M} &\equiv \int d\xv_1 d\xv_2\,\bar\psi(\xv_1)\,\Phi(\xv_1-\xv_2)W_\kappa[\Av_L;\xv_1,\xv_2] \psi(\xv_2)\ket{0}
\end{align}
At \order{\as^0}, \ie, neglecting the \order{g} interactions in the Hamiltonian \eq{4.3}, the condition $H_{QCD}\ket{q\qb;M} = M\ket{q\qb;M}$ defines the  Bound State Equation (BSE) for the wave function $\Phi(\xv)$, 
\begin{align} \label{5.11}
i\nv\cdot\acomb{\alv}{\Phi(\xv)}+m\comb{\gz}{\Phi(\xv)} &= \big[M-V(\xv)\big]\Phi(\xv)
\end{align}
where $V(\xv) = \La^2|\xv|$. Defining the kinetic energies of the quark $\rham{0}$ and antiquark $\lham{0}$ by
\begin{align} \label{5.12}
\rham{0} &\equiv i\rnab\cdot\alv +\gz m \hspace{3cm}
\lham{0} \equiv -i\lnab\cdot\alv +\gz m \nn\crt
\rham{0}^2 &= -\rnab^2+m^2 \hspace{3.4cm} \lham{0}^2 = -\lnab^2+m^2
\end{align}
the BSE \eq{5.11} may be expressed as
\begin{align} \label{5.13}
\rham{0}\Phi(\xv) - \Phi(\xv)\lham{0} = \big[M-V(\xv)\big]\Phi(\xv)
\end{align}
Multiplying this by $\rham{0}$ from the left and by $\lham{0}$ from the right and adding them gives,
\begin{align} \label{5.14}
\rham{0}^2\Phi(\xv)-\Phi(\xv)\lham{0}^2 = \rham{0}\big[(M-V)\Phi(\xv)\big] + \big[(M-V)\Phi(\xv)\big]\lham{0}
\end{align}
Since $\rham{0}^2$ and $\lham{0}^2$ are Dirac scalars \eq{5.12} the BSE implies
\begin{align} \label{5.15}
\rham{0}\big[(M-V)\Phi(\xv)\big] + \big[(M-V)\Phi(\xv)\big]\lham{0} = 0
\end{align} 

Some properties of the wave function $\Phi(\xv)$ were studied in \cite{Hoyer:2021adf}. The radial and angular variables can be separated in the BSE by expanding $\Phi(\xv)$ in 16 Dirac structures. The states are classified by their angular momentum~$j$, parity $\eta_P=\pm 1$ and charge conjugation $\eta_C = \pm 1$. There are no solutions of \eq{5.11} with $\eta_P=-\eta_C = (-1)^j$, which would be exotic in the non-relativistic quark model. For small quark masses $m$ the states lie on linear Regge trajectories, $j \simeq \alpha_0 +\alpha' M^2$, with parallel daughter trajectories.

At large quark separations the \order{\as^0} wave function oscillates, $\Phi(\xv\to\infty) \sim \exp(\pm i\xv^2\La^2/4)$. The BSE is satisfied by negative quark kinetic energies canceling the large potential. This is  similar to the Dirac equation with a linear potential, where the oscillating wave function describes positrons repulsed by the potential \cite{Hoyer:2021adf}. Here \order{\as} $q\qb$ pair production (``string breaking'') is important at large values of the potential. The \order{\as^0} oscillations with constant norm may reflect quark pair production in an average (dual) sense. This remains to be studied.

\section{Chiral symmetry} \label{secVI}

As previously noted, the solutions $\Phi(\xv)$ of the BSE \eq{5.11} are characterized by their parity $\eta_P=\pm 1$, charge conjugation $\eta_C=\pm 1$ and angular momentum $j$ \cite{Hoyer:2021adf}. With $T$ denoting the transpose, 
\begin{align} \label{6.1}
\gz\Phi(-\xv)\gz = \eta_P\Phi(\xv) \hspace{3cm} \alpha_2[\Phi(-\xv)]^T\alpha_2 = \eta_C\Phi(\xv)
\end{align}

For vanishing quark mass $m=0$ there are parity doublets: If $\Phi(\xv)$ solves \eq{5.11} then so does $\gf\Phi(\xv)$, with opposite $\eta_P$. This reflects the chiral symmetry of the QCD action \eq{4.1} under global transformations $\psi(x) \to \exp(i\vphi\gf)\psi(x)$. Chiral SU(2) symmetry is well satisfied in the light quark sector of QCD, since the $u$ and $d$ quark masses are small compared to the confinement scale, $m_{u,d} \ll \La_{QCD}$. Yet hadrons lack parity doublets. The chiral symmetry of the action is apparently spontaneously broken, \ie, the ground state is not chiral invariant. The consequences of spontaneous chiral symmetry breaking (CSB) are in good agreement with data \cite{Weinberg:1996kr,Scherer:2002tk,Nefediev:2025zkv}. 

Large mass splittings between opposite parity hadrons are found in lattice QCD calculations even in the quenched approximation, which neglects quark loops \cite{Fodor:2012gf}. CSB may thus appear already at \order{\as^0} in the present framework. This requires a quark condensate $\bra{0}q\bar q\ket{0} \neq 0$, which serves an order parameter of CSB \cite{GellMann:1968rz}. In the following I study how some aspects of CSB might be realized.

The BSE \eq{5.11} has for $m=0$ a solution $\Phi_\sigma(\xv)$, which defines a massless ($M_\sigma =0$) $j^{PC}=0^{++}$ state $\hat\sigma\ket{0}$ \cite{Hoyer:2021adf},
\begin{align} \label{6.2}
&\ket{\sigma} \equiv \hat\sigma\ket{0} \equiv \int d\xv_1 d\xv_2\,\qb(\xv_1)\,\Phi_\sigma(\xv_1-\xv_2)W_\kappa[\Av_L]  \,q(\xv_2)\ket{0} \hspace{2cm} q = \left(\begin{array}{c} u \crt d \end{array} \right) \nn\crt
&i\nv\cdot\acom{\alv}{\Phi_\sigma(\xv)}+V'|\xv|\,\Phi_\sigma(\xv)=0 \hspace{2cm}
\Phi_\sigma(\xv) = N_\sigma\big[J_0(\quart V'\xv^2)+ \frac{i\,\alv\cdot\xv}{|\xv|}\,J_1(\quart V'\xv^2)\big] 
\end{align}
where $W_\kappa[\Av_L] \equiv W_\kappa[\Av_L;\xv_1,\xv_2]$ is defined in \eq{5.1}, $V'=\La^2$ \eq{5.9} and $J_0,\,J_1$ are Bessel functions. The massless $u$ and $d$ quarks form an isospin doublet $q$. The isosinglet $q\qb$ state $\ket{\sigma}$ has 4-momentum $P =0$ in all frames, so it may mix with the perturbative vacuum without violating Poincar\'e invariance. I consider the toy ``$\sigma$ condensate'' $\ket{0}_\sigma$,
\begin{align} \label{6.3}
\ket{0}_\sigma &\equiv  \exp(\hat\sigma)\ket{0}
\end{align}
where $\hat\sigma$ is defined in \eq{6.2}. Interactions within $\ket{0}_\sigma$ are neglected. This vacuum has a chiral order parameter,
\begin{align} \label{6.4}
_\sigma{\hspace{-.5mm}}\bra{0}\bar q(x)q(x)\ket{0}_\sigma = 2\,_\sigma{\hspace{-.5mm}}\bra{0}\tr[\gz\Phi_\sigma(0)\gz]\ket{0}_\sigma = 8N_\sigma\, {_\sigma{\hspace{-.5mm}}\bra{0}}0\rangle_\sigma
\end{align}
Chiral transformations with isospin $I=1$ are generated by 
\begin{align} \label{6.5}
Q_{5}^i = \int d\xv\, q^\dag(\xv)\gf\halft\tau^i q(\xv)
\end{align}
where the $\tau^i$ are Pauli isospin matrices. They transform $\hat\sigma$ into the Goldstone boson $\hat\pi$,
\begin{align} \label{6.6}
i\com{Q_{5}^i}{\hat\sigma} &= \hat\pi^i \hspace{2.7cm} \hat\pi^i = \int d\xv_1 d\xv_2\,\bar q(\xv_1)\Phi_{\pi}^i(\xv_1-\xv_2)W_\kappa[\Av_L]\, q(\xv_2) \crt
i\com{Q_{5}^i}{\hat\pi^j} &= -\delta_{ij}\,\hat\sigma \hspace{2cm} \Phi_{\pi}^i(\xv) = -i\tau^i\gf\Phi_\sigma(\xv) \nn
\end{align}

Meson states like \eq{5.10} with $m=0$ and mass $\mM$ can be built on the $\ket{0}_\sigma$ ground state,
\begin{align} \label{6.7}
\ketb{\Phi_\mM}_\sigma &\equiv \int d\xv_1 d\xv_2\,\qb(\xv_1)\Phi_\mM(\xv_1-\xv_2)W_\kappa[\Av_L] q(\xv_2)\ket{0}_\sigma \equiv \hat\mM\ket{0}_\sigma
\end{align}
In the absence of interactions between $\hat\mM$ and $\hat\sigma$ the \order{\alpha_s^0} eigenstate condition $(H-\mM)\ket{\Phi_\mM}_\sigma = 0$ \eq{5.11} implies,
\begin{align} \label{6.8}
i\rnab_x\cdot\acomb{\alv}{\Phi_\mM(\xv)} =(\mM-V(\xv))\Phi_{\mM}(\xv)
\end{align}
where $V(\xv)=V'|\xv|$. The condensate vacuum $\ket{0}_\sigma$ can influence the bound state $\ketb{\Phi_\mM}_\sigma$  through quark mixing. Including in \eq{6.7} a contraction of $\bar q(\xv_1)$ in $\hat\mM$ with $q(\xv_2)$ in $\hat\sigma$ \eq{6.2}, and \textit{vice versa}, changes the state by
\begin{align}
\ketb{\delta\Phi_\mM}_\sigma &= \int d\xv_1 d\xv_2\,\qb(\xv_1)\delta\Phi_\mM(\xv_1-\xv_2) W_\kappa[\Av_L] q(\xv_2)\ket{0}_\sigma  \label{6.9}\crt
\delta{\Phi}_\mM(\xv) &= \int d\yv\big[\Phi_\mM(\xv-\yv)\gz\Phi_\sigma(\yv) - \Phi_\sigma(\yv)\gz\Phi_\mM(\xv-\yv)\big] \label{6.10}
\end{align}
Treating the correction as a perturbation allows to estimate the change of bound state mass $\delta\mM$ as
\begin{align} \label{6.11}
\delta\mM = \frac{(_\sigma\brab{\Phi_\mM}+\,_\sigma\brab{\delta\Phi_\mM)})\,H\,(\ketb{\Phi_\mM}_\sigma+\ketb{\delta\Phi_\mM}_\sigma)}{_\sigma\brab{\Phi_\mM}\Phi_\mM\rangle_\sigma}-\mM \ 
\simeq \  \frac{_\sigma\brab{\Phi_\mM}\delta\Phi_\mM\rangle_\sigma+\,_\sigma\brab{\delta\Phi_\mM)}\Phi_\mM\rangle_\sigma)}{_\sigma\brab{\Phi_\mM}\Phi_\mM\rangle_\sigma}\,\mM
\end{align}

The parity and charge conjugation of $\ketb{\delta\Phi_\mM}_\sigma$ \eq{6.9} should agree with $\ketb{\Phi_\mM}_\sigma$ for $_\sigma\brab{\Phi_\mM}\delta\Phi_\mM\rangle_\sigma$ not to vanish. Let us verify that \eq{6.1} for $\Phi_\mM(\xv)$ holds also for $\delta\Phi_\mM(\xv)$, given that $\Phi_\sigma(\xv)$ satisfies \eq{6.1} with $\eta_P^\sigma=\eta_C^\sigma = +1$.

\vspace{-.5cm}

\begin{align}
&\textit{Parity}:\hspace{.5cm} \gz\delta\Phi_{\mM}(-\xv)\gz = \int d\yv\big[\eta_P^\mM\Phi_\mM(\xv+\yv)  
\gz\Phi_\sigma(-\yv)-\Phi_\sigma(-\yv)\gz\eta_P^\mM\Phi_\mM(\xv+\yv)\big] =\eta_P^\mM \delta\Phi_{\mM}(\xv)\label{6.12}\crt
&\textit{Charge conjugation:}\hspace{.5cm} \alpha_2\delta\Phi_{\mM}^T(-\xv)\alpha_2 = \int d\yv\,\alpha_2\big[\Phi_\sigma^T(\yv)\gz\Phi_\mM^T(-\xv-\yv)-\Phi_\mM^T(-\xv-\yv)\gz\Phi_\sigma^T(\yv)\big]\alpha_2 \nn\crt
& \hspace{3.5cm}=\eta_C^\mM \int d\yv\big[\Phi_\sigma(-\yv)\alpha_2\gz\alpha_2\Phi_\mM(\xv+\yv)-\Phi_\mM(\xv+\yv)\alpha_2\gz\alpha_2\Phi_\sigma(-\yv)\big] = \eta_C^\mM \delta\Phi_{\mM}(\xv) \label{6.13}
\end{align}

The \textit{angular momentum} of $\ketb{\delta\Phi_\mM}_\sigma$ also agrees with that of $\ketb{\Phi_\mM}_\sigma$. The generator $\bs{\mJ}$ of quark field rotations satisfies
\begin{align} \label{6.14}
\bs{\mJ} &= \int d\xv\,q^\dag(\xv)\,\bs{J}\,q(\xv) \hspace{2.9cm} \bs{J} = \Lv+\Sv= \xv\times(-i\rnab)+\halft\gf\alv \nn\crt
\bs{\mJ}\ket{\Phi_\mM}_\sigma&=\int d\xv_1 d\xv_2\, \bar q(\xv_1) \com{\Jv}{\Phi_\mM(\xv)}W_\kappa[\Av_L]\,q(\xv_2)\ket{0}_\sigma
\end{align}
From $\com{\Jv}{\Phi_{\sigma}(\xv)}=\com{\Jv}{\gz}=0$ follows that $\com{\Jv}{\delta\Phi_{\mM}(\xv)}$ transforms as $\com{\Jv}{\Phi_\mM(\xv)}$. Hence the $j^{PC}$ quantum numbers of the state are not affected by the CSB. 

On the other hand, the CSB does remove the mass degeneracy of the states $\ket{\Phi_\mM}_\sigma$ and $\ket{\gf\Phi_\mM}_\sigma$. In \eq{6.11},
\begin{align} \label{6.15}
_\sigma\brab{\Phi_\mM}\delta\Phi_\mM\rangle_\sigma &\propto \int d\xv\,\tr\big\{\Phi_\mM^\dag(\xv)\delta\Phi_\mM(\xv)\big\} = \int d\xv\,\tr\big\{\Phi_\mM^\dag(\xv)\gf\,\gf\delta\Phi_\mM(\xv)\big\} \nn\crt
\gf\delta\Phi_\mM(\xv) &= \int d\yv \big[\gf\Phi_\mM(\xv)\gz\Phi_\sigma(\yv) + \Phi_\sigma(\yv)\gz\gf\Phi_\mM(\xv)\big]
\end{align}
The last term has opposite sign compared to the overlap for the parity partner state $\ket{\gf\Phi_\mM}_\sigma$, removing the degeneracy of the parity doublets. The general features of the \order{\as^0} states built on the $\ket{0}_\sigma$ vacuum \eq{6.3} thus agree with CSB.

\section{Discussion} \label{secVII}

QCD hadrons are strongly bound, apparently requiring a large (non-perturbative) value of $\as$. The ``apparently'' is motivated by the successful phenomenology \cite{Eichten:1979ms,Eichten:2007qx} of heavy quarkonia, in terms of the Schr\"odinger equation with the Cornell potential \eq{1.3}. The confining potential $V'r$ provides strong binding, allowing the QCD coupling to be moderate, $\as \simeq 0.39$. Quarkonium decays and spin dependence are evaluated perturbatively in powers of $\as$, as for Positronium in QED. 

Confinement determines the radii of hadrons and thus the minimum scale in $\as(Q^2)$. If the coupling freezes at a moderate value \cite{Gribov:1999ui,Dokshitzer:2004ie} a perturbative expansion can be meaningful even for hadrons made of light quarks. This is supported by the successful (quark model) classification of hadrons in terms of their valence quarks. Fock states with multiple quark pairs and gluons also contribute, but are suppressed by powers of $\as$.

Gauge theories have instantaneous interactions, which can bind even relativistic constituents without retardation. The temporal gauge condition $A^0(t,\xv) =0$ is imposed over all space at an instant of time. Physical states are constrained by Gauss' law, which requires charged constituents to carry a longitudinal gauge field $\Av_L$ \cite{Dirac:1955uv}. This implies an electric field $\Ev_L$, whose energy contributes an instantaneous potential (the Coulomb potential of Positronium).

Gauss' law determines $\Ev_L$ only up to a boundary condition. Space translation and rotation symmetry for $q\qb$ states restricts the homogeneous contribution to the term $\propto \kappa$ in \eq{5.1}. This gives (for any quark mass) the Cornell potential \eq{5.9}, where $\La$ is a universal scale. The boundary condition which defines temporal gauge thus brings the hadron scale, which is not part of the QCD action. Standard perturbation theory derives Feynman diagrams by implicitly assuming the gauge dependent fields ($A^0\,,\Av_L$) to vanish at large distance ($\kappa=0$).

Confinement through Gauss' constraint does not involve the creation of quark pairs or gluons. This allows hadrons to be dominated by their valence Fock states even though strongly bound.
Since the scale $\La$ is unrelated to $\as$ we may consider a perturbative expansion for QCD bound states, with $\as=0$ defining the lowest order. Higher orders of $\as$ involve Fock states with propagating (transverse) gluons and additional $q\qb$ pairs, as indicated for QED in \fig{f1}. Poincar\'e covariance is ensured by the Lie algebra of the generators, which are determined by the action. An example is provided by $D=1+1$ dimensional QCD in the 't Hooft limit \cite{Hoyer:2025cpf}.

For vanishing quark mass there is a massless $j^{PC}=0^{++}$ ``$\sigma$'' eigenstate of the Hamiltonian \eq{6.2}, which may mix with the perturbative vacuum \eq{6.3} without violating Poincar\'e invariance. There is also a massless $j^{PC}=0^{-+}$ ``$\pi$'' \eq{6.6}, identified as the Goldstone boson. The mixing of bound state quarks with those in the $\sigma$ vacuum removes the parity doublet degeneracy. Further study is needed to judge whether this is a viable model for spontaneous Chiral Symmetry Breaking.

\begin{acknowledgments}
I thank Alexey Nefediev, Stephane Peign\'e and Roman Zwicky for valuable discussions. I am privileged to be associated as Professor Emeritus to the Physics Department of the University of Helsinki.
\end{acknowledgments}

\bibliography{refs.bib}
\end{document}